\documentclass[12pt]{article}
\usepackage{epsfig}
\usepackage{amsfonts}
\usepackage{mathtools} 
\usepackage{amscd}
\usepackage{latexsym}
\usepackage{amsmath,amssymb,amsthm}
\usepackage{verbatim}
\usepackage{setspace}
\usepackage{cite}
\usepackage{graphicx}
\usepackage{mathtools}
\usepackage[all]{xy}
\usepackage{tikz}
\usepackage[textheight=9in, textwidth=6.5in, letterpaper]{geometry}
\usepackage{color}   
\usepackage{hyperref}
\hypersetup{
    colorlinks=true,  
    linktoc=all,     
    linkcolor=black,  
    citecolor=black,
    filecolor=black,
    urlcolor=black,
}

\numberwithin{equation}{section}

\newtheorem{Theorem}{Theorem}[section]
\newtheorem*{Theorem*}{Theorem}

\newtheorem{Lemma}[Theorem]{Lemma}
\newtheorem{Proposition}[Theorem]{Proposition}

 { \theoremstyle{definition}
\newtheorem{Definition}[Theorem]{Definition}

\newtheorem{Example}[Theorem]{Example}
 }
\def\p{\partial}

\def\<{\langle}
\def\>{\rangle}

\def\cO{\mathcal{O}}

\def\be{\begin{equation}}
\def\ee{\end{equation}}
\def\beq{\be\begin{array}{c}}
\def\eeq{\end{array}\ee}
\def\bes{\be\begin{split}}
\def\ees{\end{split} \ee}
\def\bs{\begin{split}}
\def\es{\end{split} }
\def\nn{\nonumber}

\def\b{{\beta}}
\def\a{{\alpha}}
\def\g{{ \gamma}}

\def\e{{\epsilon}}
\def\ve{{\varepsilon}}

\def\vp{\varphi}

   \makeatletter
  \let\over=\@@over \let\overwithdelims=\@@overwithdelims
  \let\atop=\@@atop \let\atopwithdelims=\@@atopwithdelims
  \let\above=\@@above \let\abovewithdelims=\@@abovewithdelims
\renewcommand\section{\@startsection {section}{1}{\z@}%
                                   {-3.5ex \@plus -1ex \@minus -.2ex}
                                   {2.3ex \@plus.2ex}%
                                   {\normalfont\large\bfseries}}

\renewcommand\subsection{\@startsection{subsection}{2}{\z@}%
                                     {-3.25ex\@plus -1ex \@minus -.2ex}%
                                     {1.5ex \@plus .2ex}%
                                     {\normalfont\bfseries}}

\begin{document}
\begin{titlepage}
\unitlength = 1mm

\vskip 1cm
\begin{center}
{ \LARGE {\textsc{Genus-one Landau-Ginzburg-Saito theory for descendant Gromov-Witten theory on the projective line}}}

\vspace{1cm}
 Vyacheslav Lysov\\
\vspace{1cm}
{\it Center for Mathematics and Interdisciplinary Sciences,\\ Fudan University, Shanghai, 200433, China}\\
\vspace{0.3cm}
{\it Shanghai Institute for Mathematics and Interdisciplinary Sciences (SIMIS),\\ Shanghai, 200433, China}\\
\vspace{0.3cm}

\vspace{0.8cm}
\vspace{0.8cm}

\begin{abstract}
We define the genus-one correlation functions for the descendants in the Landau-Ginzburg-Saito theory. We show that the correlation functions obey puncture, divisor, dilaton, and genus-one topological recursion relations. We define a mirror map for the genus-one observables. We prove that the genus-one LGS correlation functions of the mirror observables are equal to the genus-one GW invariants on the projective line with descendants. 

\end{abstract}
\end{center}
\end{titlepage}

\pagestyle{empty}
\pagestyle{plain}
\pagenumbering{arabic}
\tableofcontents

\section{Introduction}

Mirror symmetry in mathematical physics is a map between correlation functions in the A-model, representing Gromov-Witten (GW) invariants, and correlators of mirrored observables in the B-model, representing complex-structure deformations. The most studied case of mirror symmetry maps GW theory for a toric Fano variety to the Landau-Ginzburg (LG) theory on a complex torus with a mirror superpotential. In this setting, mirror symmetry relates complicated integrals over the moduli space of holomorphic maps in GW theory to simpler multidimensional residue integrals in LG theory. 

The simplest example of a toric Fano variety is the projective line $\mathbb{P}^1$. The GW invariants for $\mathbb{P}^1$ are known in explicit form \cite{dubrovin2019gromov} in full generality. Namely, they include gravitational descendants at arbitrary genus. Moreover, we know many ways to evaluate them, including the topological recursion relations, the Toda conjecture \cite{pandharipande2000toda}, the Hurwitz numbers relation \cite{okounkov2006gromov}, integrability \cite{dubrovin2019gromov}, and many others. The mirror LG model has target space $\mathbb{C}^{\ast}$, holomorphic top form $\Omega = dY$ and simple mirror superpotential $W = e^{iY} + q e^{-iY}$. Hence, we expect a simple B-model evaluation of the invariants in the same generality via the residue-type integrals. 

Unfortunately, no such B-model construction in full generality is available in the literature. We only know the partial results. The mirror map for $\mathbb{P}^1$ is extremely easy due to the simplicity of the GW invariants on $\mathbb{P}^1$. The higher-genus invariants without gravitational descendants vanish, while the genus-zero invariants are given by the residue formula for the deformed superpotential $W = t_0 + e^ {iY} + q e^{-iY}e^{t_1}$. The nontrivial structure of the deformation, the $e^{t_1}$ factor, originates from the K. Saito theory of good sections \cite{saito1983period}. Hence, we refer to an LG model with good-section data as the Landau-Ginzburg-Saito (LGS) model to indicate the importance of good-section data. In our previous work \cite{Lysov:2025nmn}, we defined genus-zero LGS correlation functions for B-model descendants. We used the Kontsevich-Manin map \cite{kontsevich1998relations} for gravitational descendants to prove the genus-zero mirror relation for $\mathbb{P}^1$ with gravitational descendants. 

In this paper, we present a first step toward higher-genus mirror symmetry for $\mathbb{P}^1$: the genus-one mirror relation. The genus-one GW invariants admit an algebraic description via the genus-one topological recursion and the puncture and divisor relations. Hence, we can use the same strategy as in \cite{Lysov:2025nmn} to prove the genus-one mirror relation. To do so, we first generalize our definition of the LGS theory to include genus-one correlation functions. The Kontsevich-Manin map \cite{kontsevich1998relations} is genus-independent, so we do not need to modify it for genus one. 

The paper is organized as follows. Section 1 reviews the properties of GW invariants at genera zero and one for $\mathbb{P}^1$. Section 2 defines the LGS correlation functions for descendants and discusses the corresponding puncture, divisor, dilaton, and topological recursion relations. In Section 3, we formulate the mirror map relation between the GW descendants and LGS descendants, and formulate and prove our main theorem on the mirror relation between the GW invariants and LGS correlation functions with descendants. In Section 4, we demonstrate the mirror theorem for several well-known examples of the GW invariants and prove polynomiality for the genus-one descendant GW invariants.

\section{Gromov-Witten invariants}

In this section, we briefly review the definition and properties of GW invariants, and further specify the genus-one case for $\mathbb{P}^1$.
\begin{Definition}
    For a collection of $ \gamma_\alpha \in H_{dR} (X)$ on a projective toric variety $X$, integers $ m_\alpha \in \mathbb{Z}_{\geq 0}$ and $\beta \in H_2(X)$, the GW invariant is the integral over the (properly completed) moduli space of genus-$g$ holomorphic maps $\mathcal{M}_{g,n}(X,\beta)$
    \be\label{eq_GW_desc_def}
        \< \tau_{m_1} (\g_1)\cdots \tau_{m_n}(\g_n)\>_{g,\beta} = \int_{\overline{ \mathcal{M}_{g,n}(X,\beta)}} \bigwedge_{\a=1}^n \;\psi_\a^{m_\a}\;ev_\a^\ast \g_\a.
    \ee
\end{Definition}
For $m=0$ we will often use a simplified notation $\g$ instead of $\tau_0(\g)$. We introduce the generating function for invariants of different degrees
\be
\begin{split}
    \< \tau_{m_1} (\g_1)\cdots \tau_{m_n}(\g_n)\>_{g}  &= \sum_{\beta} q^\beta\; \< \tau_{m_1} (\g_1)\cdots \tau_{m_n}(\g_n)\>_{g,\beta}.
\end{split}
\ee
The GW invariant (\ref{eq_GW_desc_def}) vanishes unless the moduli space dimension matches the degree of the integrand. Namely,
\be
    \dim \overline{\mathcal{M}_{g,n}(X,\beta)} =  (3-\dim X) (g-1) +\int_\beta c_1(T_X) +n = \sum_{\a=1}^n \left(m_\a+\frac12 \deg \g_\a \right).
\ee
For $X =\mathbb{P}^1$ we have $\dim H_2(X) =1$, so the degree of the map is a single natural number $d$, while the moduli space dimension
\be
    \dim \overline{\mathcal{M}_{g,n}(\mathbb{P}^1,d)}= 2(g-1)+ \int_\beta c_1(T_{\mathbb{P}^1}) +n =2g-2+2d +n.
\ee
The observables $\g_\a$ on $\mathbb{P}^1$ are either point class $P\in H^2(\mathbb{P}^1)$ with Fubini-study representative $\g_P=\omega_{FS}$ or an identity class $I \in H^0(\mathbb{P}^1)$ with constant representative $\g_I=1$.

\subsection{Relations}
Here, we will specify the various relations for the GW invariants to the case of $X = \mathbb{P}^1$.
\begin{Theorem} {\bf (Puncture relation)} 
    For either $n+2g\geq 3$ or $d>1, n\geq 1$
    \be\label{eq_GW_puncture}
    \begin{split}
        \<I\;& \tau_{m_1}(\g_1)\cdots \tau_{m_n}(\g_n)\>_{g,d}= \sum_{i=1}^n \<  \tau_{m_1}(\g_1)\cdots \tau_{m_{i-1}}(\g_{i-1})\tau_{m_i-1}(\g_i)\tau_{m_{i+1}}(\g_{i+1}) \cdots \tau_{m_n}(\g_i)\>_{g,d}.
    \end{split}
\ee 
    There is a special case $n=3, g=0$ and $d=0$
    \be
        \<I\;\gamma_1\;\gamma_2\>_{0,0} = \int_{\mathbb{P}^1} \gamma_1\wedge \gamma_2.
    \ee
\end{Theorem}
\begin{Theorem}{\bf(Dilaton relation)}
    For either $n+2g\geq 3$ or $d>1, n\geq 1$ 
    \be\label{eq_GW_dilaton}
        \<\tau_{1}(I)\tau_{m_1}(\g_1)\cdots \tau_{m_n}(\g_n)\>_{g,d} = (2g-2+n)\< \tau_{m_1}(\g_1) \cdots \tau_{m_n}(\g_1)\>_{g,d}.
    \ee
    There is a special case for $g=1, d=0$
    \be
        \<\tau_1(I)\>_{1,0} = \frac{\chi(\mathbb{P}^1)}{24} = \frac{1}{12}.
    \ee
\end{Theorem}
\begin{Theorem} {\bf (Divisor relation)} 
    For $d>0$ or $d=0,$ $2g-2+n>0$
    \be\label{eq_GW_divisor}
    \begin{split}
        \<P\;&\tau_{m_1}(\g_1)\cdots\tau_{m_n}(\g_n)\>_{g,d}=d \cdot \< \tau_{m_1}(\g_n)\cdots\tau_{m_n}(\g_n)\>_{g,d} \\
        &+ \sum_{i=1}^n \<\tau_{m_1}(\g_1)\cdots  \cdot \tau_{m_{i-1}}(\g_{i-1}) \cdot \tau_{m_i-1} (\g_i \wedge \gamma_P) \cdot\tau_{m_{i+1}}(\g_{i+1})\cdots \tau_{m_n} (\g_n)\>_{g,d}.
    \end{split}
    \ee 
    There are special cases 
    \be
        \<P\;\gamma_1\;\g_2\>_{0,0}  = \int_{\mathbb{P}^1} \gamma_P\wedge\gamma_1\wedge \g_2.
    \ee
    \be
        \<\tau_0(P)\>_{1,0} =  - \frac{1}{24}.
    \ee
\end{Theorem}

\subsection{Topological recursion relations}
The intersection matrix for $H^\ast (\mathbb{P}^1)$
\be\label{eq_intersection_matrix}
    \eta_{II} = \int_{\mathbb{P}^1} \g_I\wedge \g_I = 0,\;\;\; \eta_{PI} = \int_{\mathbb{P}^1} \g_P\wedge \g_I = 1,\;\;\; \eta_{PP} = \int_{\mathbb{P}^1} \g_P\wedge \g_P = 0.
\ee
The corresponding inverse matrix
\be
    \eta^{II} = \eta^{PP}=0,\;\;\; \eta^{IP} = \eta^{PI}=1.
\ee
\begin{Theorem} {\bf (Genus-zero topological recursion relation, Witten\cite{witten1990two})}
    For $n\geq 3$ and $m_1\geq 1$
    \be\label{eq_GW_trr}
    \begin{split}
        \< \tau_{m_1} (\gamma_1) &\tau_{m_2}(\gamma_2) \tau_{m_3}(\gamma_3) \prod_{i\in S} \tau_{m_i} (\gamma_i)\>_0 \\
        & = \sum_{S_1\sqcup S_2} \left\<I\; \tau_{m_1-1} (\gamma_1)\;  \prod_{\a\in S_1} \tau_{m_\a} (\gamma_\a)\right\>_0 \left\< P\; \tau_{m_2}(\gamma_2)\; \tau_{m_3}(\gamma_3) \; \prod_{\b\in S_2} \tau_{m_\b} (\gamma_\b)\right\>_0 \\
        & +\left\<P\;\tau_{m_1-1} (\gamma_1)\;  \prod_{\a\in S_1} \tau_{m_\a} (\gamma_\a)\right\>_0 \left\< I\; \tau_{m_2}(\gamma_2)\; \tau_{m_3}(\gamma_3) \prod_{\b\in S_2} \tau_{m_\b} (\gamma_\b)\right\>_0.
    \end{split}
    \ee
\end{Theorem}
    Note that the theorem allows for $S=\emptyset$, namely, it includes the TRR for the 3-point function 
    \be\nn
        \< \tau_{m_1} (\gamma_1)\tau_{m_2}(\gamma_2) \tau_{m_3}(\gamma_3)\> 
        = \< \tau_{m_1-1} (\gamma_1) I\> \<P \tau_{m_2}(\gamma_2) \tau_{m_3}(\gamma_3)\>+\< \tau_{m_1-1} (\gamma_1) P\> \<I \tau_{m_2}(\gamma_2) \tau_{m_3}(\gamma_3)\>.
    \ee  
\begin{Theorem}{\bf (Genus-one topological recursion relation, Witten\cite{witten1990two})}\label{TRR_genus_one} 
    For $m_1>0$ and $S = \{2,\ldots,n\}$
    \be\label{eq_GW_trr_genus_one}
    \begin{split}
        &\left\< \tau_{m_1} (\gamma_1)  \prod_{i\in S} \tau_{m_i} (\gamma_i)\right\>_1 =\frac{1}{24} \eta^{ab}\left\<\g_a \g_b\tau_{m_1-1} (\gamma_1) \prod_{i\in S} \tau_{m_i} (\gamma_i)\right\>_0 \\
        & \qquad\qquad+ \sum_{S_1\sqcup S_2} \eta^{ab}\left\<\g_a\; \tau_{m_1-1} (\gamma_1)\; \prod_{\a\in S_1} \tau_{m_\a} (\gamma_\a)\right\>_0 \left\< \g_b\; \; \prod_{\b\in S_2} \tau_{m_\b} (\gamma_\b)\right\>_1.
    \end{split}
    \ee
\end{Theorem}
\begin{Lemma}\label{lemma_GW_genus_one_no_descend}
    The GW invariants at level zero descendants at genus-one vanish unless $m=0$ and $n=1$
    \be
        \< \underbrace{P\cdots P}_{n}\underbrace{I\cdots I}_{m}\>_{1} = -\frac{1}{24} \delta_{m,0}\delta_{n,1}.
    \ee
\end{Lemma}
\begin{proof}
    If $m>0$, we use the puncture relation (\ref{eq_GW_puncture}) to show that it is zero. For $m=0$, we use the divisor relation (\ref{eq_GW_divisor}) till we reach a single observable special case
    \be
        \< \underbrace{P\cdots P}_{n}\>_1 = q\p_q\< \underbrace{P\cdots P}_{n-1}\>_1=(q\p_q)^{n-1} \<P\>_1 =(q\p_q)^{n-1}  \left(-\frac{1}{24}\right)=0.
    \ee
\end{proof}
\begin{Proposition} 
    The genus-one 1-point function for point descendants on $\mathbb{P}^1$
    \be
        \<\tau_{2n}(P)\>_{1} = \frac{2n-1}{24\; n!^2}q^n. 
    \ee
\end{Proposition}
\begin{proof}
    We will prove the relation by induction. For $n=0$, we have one of the genus one special cases (\ref{special_case_genus_one}). For $n>1$, we use the 1-point version of the genus one TRR (\ref{eq_GW_trr_genus_one}). Namely
    \be
        \<\tau_{2n}(P)\>_1 = \<\tau_{2n-1}(P) I\>_0 \cdot \<P\>_1 +\<\tau_{2n-1}(P) P\>_0 \cdot \<I\>_1+ \frac{1}{12}\cdot \<\tau_{2n-1}(P) IP \>_0. 
    \ee
    We use $\<I\>_1=0$, $\<P\>_1 =- \frac{1}{24} $, divisor and the puncture relations to simplify
    \be
        \<\tau_{2n}(P)\>_1 = -\frac{1}{24}\<\tau_{2n-2}(P)\>_0  + \frac{1}{12}\cdot \<\tau_{2n-2}(P)P \>_0 = \frac{2n-1}{24}\<\tau_{2n-2}(P)\>_0 =\frac{2n-1}{24\; n!^2}q^n  . 
    \ee
\end{proof}
\begin{Proposition} 
    The genus-one 1-point functions of the identity-class descendants
    \be
        \<\tau_{2n+1}(I)\>_{1} = \frac{2q^n}{n!^2} \left(\frac{1}{24} - \frac{2n-1}{24}H_n\right),\;\; H_n = \sum_{k=1}^n \frac{1}{k}. 
    \ee
\end{Proposition}
\begin{proof} 
    We will prove the relation by induction. For $n=0$, we have one of the genus one special cases (\ref{special_case_genus_one}). For $n>1$, we use the 1-point version of the genus one TRR (\ref{eq_GW_trr_genus_one}). Namely
    \be
        \<\tau_{2n+1}(I)\>_1 = \<\tau_{2n}(I)I\>_0 \cdot\<P\>_1 +\<\tau_{2n}(I)P\>_0 \cdot \<1\>_1+ \frac{1}{12}\cdot \<\tau_{2n}(I)IP\>_0. 
    \ee
    We use $\<I\>_1=0$, $\<P\>_1 =- \frac{1}{24} $, divisor and the puncture relations to simplify
    \be
    \begin{split}
        \<\tau_{2n+1}(I)\>_1 &= -\frac{1}{24}\<\tau_{2n-1}(I)\>_0+\frac{1}{12}\cdot\<\tau_{2n-1}(I)P\>_0 \\
        &= \frac{2n-1}{24}\<\tau_{2n-2}(I)\>_0+\frac{1}{12}\cdot \<\tau_{2n-2}(P)\>_0 =\frac{2n-1}{24\; n!^2}q^n(-2H_n)+\frac{1}{12}\frac{q^n}{n!^2} \\
        & = \frac{2q^n}{n!^2} \left(\frac{1}{24}-\frac{2n-1}{24}H_n\right).
    \end{split}
    \ee
\end{proof}
We can use puncture, divisor, and topological recursion relations to define GW invariants in purely algebraic way.
\begin{Theorem} \label{thm_reconstruction}
    The GW TRR (\ref{eq_GW_trr_genus_one}) puncture and divisor relations define all genus-one GW invariants on $\mathbb{P}^1$ in terms of the following initial data
    \begin{itemize}
        \item GW invariants (no descendants) at genus one
        \item Descendant GW invariants at genus zero
        \item Special invariants
        \be\label{special_case_genus_one}
            \<\tau_1(I)\>_1 = \frac{1}{12},\;\;\; \<\tau_0(P)\>_{1,0} =  - \frac{1}{24}.
        \ee
    \end{itemize}
\end{Theorem}
\begin{proof}
    Suppose we start with the GW invariant $\<\tau_{m_1}(\g_1)\cdots \tau_{m_n}(\g_n)\>_1$. If the total descendant level is zero, we use Lemma \ref{lemma_GW_genus_one_no_descend}. For positive total descendant level, we use puncture and divisor relations till we eliminate all observables with zero descendant level. For the $n$-point function with all positive descendant levels, we use the TRR relation (\ref{eq_GW_trr_genus_one}) to express the $n$-point invariant in terms of invariants with fewer points. We repeat the whole procedure till the number of points reaches one. We have already evaluated the one-point genus-one GW invariants from the topological recursion relation.
\end{proof}
We can use the TRR to obtain explicit formulas for the higher-point functions. Below we present the final results for the 2- and 3- point functions from Norbury-Scott \cite{norbury2014gromov}.
\be\label{eq_GW_2pt_NS}
\begin{split}
    \<\tau_{2n}(P) \tau_{2m}(P)\>_1 &=\frac{2n^2+2m^2+2mn-n-m}{24\; n!^2\; m!^2},\\
    \<\tau_{2n-1}(P) \tau_{2m-1}(P)\>_1 &= \frac{mn}{24\; n!^2\; m!^2} (2n^2+2m^2+2mn-3m-3n).
\end{split}
\ee
The 3-point functions
\be
\begin{split}
    \<\tau_{2n_1}(P) \tau_{2n_1}(P) \tau_{2n_3}(P)\>_1&=\frac{1}{24\prod n_i!^2}\left(4 n_1n_2n_3+ \sum_{i=1}^3 n_i^2(2n_i-1) + \sum_{i\neq j} n_in_j (4n_i-1)\right),\\
    \<\tau_{2n_1}(P) \tau_{2n_1-1}(P) \tau_{2n_3-1}(P)\>_1 &=\frac{n_2n_3}{24\prod n_i!^2}\left(4n_1n_2n_3+2n_1^2-3n_1-2n_2n_3 \right.\\
    &+ \left.\sum_{i=1}^3 2n_i^3-5n_i^2+3n_i + \sum_{i\neq j} n_in_j (4n_i-3)\right).
\end{split}
\ee

\subsection{GW invariants of a point}
In our definition for the LGS recursion, we will use the GW invariants for a point $X = pt$. We denote 
\be\label{eq_def_GW_point}
    \< \tau_{m_1}\cdots \tau_{m_n}\>_g = \int_{\overline{\mathcal{M}_{g,n}}} \psi_1^{m_1}\cdots \psi_n^{m_n}.
\ee
The correlation functions satisfy puncture and dilaton relations
\be
\begin{split}
    \< \tau_0\tau_{m_1}\cdots \tau_{m_n}\>_g& = \sum_{k=1}^n\< \tau_{m_1}\cdots\tau_{m_k-1}\cdots \tau_{m_n}\>_g,\\
    \< \tau_1\tau_{m_1}\cdots \tau_{m_n}\>_g& = (2g-2+n)\<\tau_{m_1}\cdots \tau_{m_n}\>_g.
\end{split}
\ee
The dilaton relation and the initial condition $\<\tau_0^3\>_0 =1$ is enough to evaluate all correlators in genus zero. Namely, 
\be
    \<\tau_{m_1}\cdots \tau_{m_n}\>_0 = \binom{n-3}{m_1, \ldots, m_n}.
\ee
Similarly, the genus one invariants are generated by the puncture and dilaton relations, by the initial condition 
\be
    \<\tau_1\>_1 = \frac{1}{24}.
\ee
\begin{Example} We can evaluate the genus-one GW invariant of a point with $n$ observables of descendant level one by a repeated application of the dilaton relation. Namely,
    \be
        \<\tau_1^n\>_1 = (n-1) \< \tau_1^{n-1}\>_1 =\ldots = (n-1)! \<\tau_1\>  = \frac{(n-1)!}{24}.
    \ee
\end{Example}
The full formula can be derived for the genus-one GW invariants of the point
\be
    \<\tau_{m_1}\cdots \tau_{m_n}\>_1 = \frac{1}{24} \binom{n}{m_1,\ldots, m_n}\left(1 - \sum_{k=2}^n \frac{(k-2)!(n-k)!}{n!}e_k(m_1,\ldots, m_n)\right).
\ee
Here, $e_k$ is the $k$'th symmetric function 
\be
    e_k(m_1, \ldots, m_n) = \sum_{i_1<\ldots<i_k} m_{i_1}\cdots m_{i_k}.
\ee

\section{LGS at genus-one}

\subsection{Definitions}
We generalize our definition \cite{Lysov:2025nmn} for the LGS correlation functions at genus zero to include the $g=1$ case.
\begin{Definition} 
    For a collection of good section $\vp_{a_k}$ and descendant levels $m_k\in \mathbb{Z}^{\geq 0}$ the LGS correlation function $\<z^{m_1}\vp_1,\ldots, z^{m_n}\vp_n\>_W^{g}$ for $g=0, n\geq 3$ and $g=1, n\geq 1$
    \begin{itemize}
        \item Vanishes for the over-extremal case $\sum m_k> 3g-3+n$
        \be
            \<z^{m_1}\vp_1,\ldots, z^{m_n}\vp_n\>_W^{g} = 0.
        \ee
        \item In the extremal case $\sum m_k =3g-3+n$ has a structure
        \be\label{def_LGS_extremal}
            \<z^{m_1}\vp_1,\ldots, z^{m_n}\vp_n\>_W^{g} = \<\tau_{m_1}\cdots\tau_{m_n}\>_{g} \sum_{p\in Crit(W)} (W'')^{g-1}\vp_1\cdots \vp_n (p).
        \ee
        \item The special under-extremal case: $g=1,\; n=1$
        \be\label{eq_def_LGS_special}
             \< \vp_P\>^1_W = -\frac{1}{24},\;\; \<\vp_I\>^1_W = 0.
        \ee
        \item For the under-extremal case $\sum m_k < 3g-3+n$, we have at least one level-zero descent level, which we relabel as $\vp$.
        The correlation function is given by the LGS recursion formula with respect to $\vp$.  
        \be\label{eq_def_LGS_recursion}
        \begin{split}
            \<z^{m_1}\varphi_1,&\ldots,  z^{m_{n-1}}\varphi_{n-1},\varphi_{n}\>^{g}_{W}=\frac{d}{d\ve}\Big|_{\ve=0}\< z^{m_1}\varphi^\ve_1,z^{m_2}\varphi^\ve_2,\ldots,z^{m_{n-1}}\varphi^\ve_{n-1}\>^g_{W^\ve}.
        \end{split}
        \ee
        The contact terms are the same as in the genus-zero case
        \be\label{eq_contact_term_desc_expansion}
        \begin{split}
            W^\ve &= W + \ve \vp_n,\\
            z^{m}\varphi^\ve &=z^m \vp +\ve C_W(z^m\vp, \vp_n)=z^m \vp+ \ve z^{m-1}S_W \pi_W(\varphi \varphi_n) +\ve z^m C_W(\varphi, \varphi_n).
        \end{split}
        \ee
        \item The deformation of a superpotential by a good section for a point observable is identical to the rescaling of the Kahler module of $W$, i.e.
        \be
            W (Y; q) + \ve \varphi_P =W (Y; q) +\ve   qe^{-iY} = W(Y; q(1+\ve)).
        \ee
        \item For the mirror of $\mathbb{P}^1$ the contact terms in (\ref{eq_contact_term_desc_expansion}) simplify to 
        \be
        \begin{split}
            C_{W}(z^{m}\vp_I, \vp_I) &= z^{m-1}\vp_I,\;\; C_{W}(z^{m}\vp_P, \vp_I) = z^{m-1} \vp_P,\\
            C_{W}(z^{m}\vp_I, \vp_P) &= z^{m-1}\vp_P,\; C_{W}(z^{m}\vp_P, \vp_P) = z^{m-1}q \vp_I+  z^m \vp_P.
        \end{split}
        \ee
    \end{itemize}   
\end{Definition}
The sum over critical points equals the residue integral
\be
    \sum_{p\in Crit(W)} (W'')^{g-1}\vp_1\cdots \vp_n (p) =\oint\frac{(W'')^g}{W'} \vp.
\ee

\begin{Example}
    The 1-point correlation functions $\<z^m\vp\>_W^1$ for $m\geq 2$ vanish. The $m=0$ correlators are given by special cases. The $m=1$ correlators are extremal. Namely
    \be
     \begin{split}
        \<z\vp_I\>_W^1 =\<\tau_1\> \cdot \oint\frac{W''}{W'} \vp_I  = 2\cdot \frac{1}{24} = \frac{1}{12}. 
        \<z\vp_P\>_W^1& = \<\tau_1\>\cdot \oint\frac{W''}{W'} \vp_P =0.
    \end{split}
    \ee
\end{Example}
\begin{Example}
    The 2-point correlation functions $\<z^{m_1}\vp_1,z^{m_1}\vp_1\>_W^1$ vanish for $m_1+m_2\geq 3$. 
    \be
        \begin{split}
        \<\vp_I, \vp_P\>^1_W& =\<\vp_I, \vp_I\>^1_W  = 0,\\
        \<\vp_I, z\vp_P\>^1_W &= \<\vp_P\>^1 = -\frac{1}{24},\;\; \<\vp_I, z \vp_I\>^1_W  =\<\vp_I\>^1 = 0.
        \end{split}
    \ee
    We use the dilaton relation 
    \be
        \<z 1, \vp_\a\>^1_W =\<\vp_\a\>^1_W,\;\; \<z 1, z\vp_\a\>^1_W =\<z\vp_\a\>^1_W.
    \ee
    The extremal correlators 
    \be
        \begin{split}
        \<z \vp_1, z\vp_2\>^1 &= \<\tau_1^2\>\cdot  N^1(\vp_1\vp_2) = \frac{1}{24}\oint \frac{W''}{W'}\vp_1\vp_2,\\
        \<z^2 \vp_1, \vp_2\>^1 &= \<\tau_2\tau_0\>\cdot  N^1(\vp_1\vp_2) = \frac{1}{24}\oint \frac{W''}{W'}\vp_1\vp_2.
        \end{split}
    \ee
\end{Example}

\subsection{Relations in LGS theory}
\begin{Proposition} {\bf (LGS puncture relation)} 
    For $g=0,1$ and $2g-2+n\geq 1$ and $m_k\geq 0$
    \be\label{eq_LGS_puncture}
        \<\vp_I,z^{m_1}\vp_1,\ldots, z^{m_n}\vp_n\>_W^{g} = \sum_{k=1}^n \<z^{m_1}\vp_1,\ldots, z^{m_k-1}\vp_k,\ldots, z^{m_n}\vp_n\>_W^{g}.
    \ee
\end{Proposition}
\begin{proof}
    The LGS recursion for $\vp_I$ and the contact term formula.
\end{proof}
\begin{Proposition}{\bf(LGS Divisor relation)} 
    For $g=0,1$ and $2g-2+n\geq 1$ and $m_k\geq 0$
    \be\label{eq_LGS_divisor}
    \begin{split}
        \<&z^{m_1}\varphi_{1},\ldots,z^{m_n}\varphi_{n},\varphi_P\>^g_W=q\frac{d}{dq}\<z^{m_1}\varphi_{1},\ldots,z^{m_n}\varphi_{n}\>^g_W \\
        &\qquad+ \sum_{i=1}^n\<z^{m_1} \vp_{1},\ldots,z^{m_i-1}S_W\pi_W(\vp_{i}\vp_P),\ldots,z^{m_n}\varphi_{n}\>^g_W .
    \end{split}
    \ee
\end{Proposition} 
\begin{proof}
    We use the LGS recursion for the $\vp_P$
    \be
    \begin{split}
        \<&z^{m_1}\varphi_{1},\ldots,z^{m_n}\varphi_{n},\varphi_P\>^g_W=\frac{d}{d\ve}\Big|_{\ve=0}\<z^{m_1}\vp_{1},\ldots,z^{m_n}\vp_{n}\>^g_{W+\ve \vp_P}\\
        &+\sum_{i=1}^n\< z^{m_1}\vp_{1},\ldots,C_W(z^{m_i}\vp_{i},\vp_P),\ldots,z^{m_n}\varphi_{n}\>^g_W = q \frac{d}{dq}\<z^{m_1}\varphi_{1},\ldots,z^{m_n}\varphi_{n}\>^g_W \\
        &\qquad+\sum_{i=1}^n\<z^{m_1}\vp_{1},\ldots,C_W(z^{m_i}\vp_{i},\vp_P)-z^{m_i}q\frac{d}{dq}\vp_{i},\ldots,z^{m_n}\varphi_{n}\>^g_W. 
    \end{split}
    \ee
    We use contact term formula (\ref{eq_contact_term_desc_expansion}) and the following equality for $\vp_I =1$ and $\vp_P = q e^{-iY}$
    \be\nn
        q\frac{d}{dq}\vp - C_W (\vp, \vp_P) = 0.
    \ee
\end{proof}
\begin{Proposition} {\bf (LGS selection rule)} 
    The $n$-point function 
    \be
        \< z^{m_1}\vp_P,\ldots, z^{m_n}\vp_P,z^{l_1}\vp_I,\ldots, z^{l_k} \vp_I\>^1_W
    \ee
    vanishes unless parity
    \be
        \Pi(\{m\},\{l\}) = \sum_{j=1}^n m_j + \sum_{r=1}^k l_r + k =0\; \mod 2. 
    \ee
\end{Proposition}
\begin{proof}
    For the extremal version $\sum m_j + \sum l_r = n+k$ the parity is 
    \be
        \Pi(\{m\},\{l\})=\sum_{j=1}^n m_j + \sum_{r=1}^k l_r + k  = n+k+k = n+2k = n \; \mod 2.   
    \ee
    \be
        \< z^{m_1}\vp_P,\ldots, z^{m_n}\vp_P,z^{l_1}\vp_I,\ldots, z^{l_k} \vp_I\>^1_W = \<\tau_{m_1}\cdots \tau_{m_n}\>_1 \sum_{p\in Crit(W)} \vp_P^{n}(p). 
    \ee
    The sum over critical points vanishes for odd $n$, so the proposition holds. For the under-extremal version we have a level-0 observable, either $\vp_I$ or $\vp_P$. In the case of $\vp_I$, we use the LGS puncture relation (\ref{eq_LGS_puncture}) to express the original correlation function with parity $\Pi$ as a sum of correlation functions with the same parity $\Pi' = \Pi$. In the case of $\vp_P$, the first term in the LGS puncture relation (\ref{eq_LGS_puncture}) has parity $\Pi' = \Pi$, and the same holds for the second term due to the relation 
    \be
        S_W \pi_W (\vp_P\vp_P) = q\vp_I,\;\;\; S_W\pi_W(\vp_I\vp_P) = \vp_P.
    \ee
\end{proof}
\begin{Proposition} {\bf (LGS dilaton relation)} 
    For $n\geq 1$ and $g=0,1$
    \be\label{eq_LGS_dilaton}
        \<z\vp_I,z^{m_1}\vp_1,\ldots, z^{m_n}\vp_n\>_W^{g} = \sum_{k=1}^n (2g-2+n) \<z^{m_1}\vp_1,\ldots, z^{m_n}\vp_n\>_W^{g}.
    \ee
\end{Proposition}
\begin{proof}
    We proved the $g=0$ case in the earlier paper \cite{Lysov:2025nmn}, so we will focus on proving the $g=1$ case. We prove the equality using induction on the number of observables $n$. The base case is $n=1$.We prove the $n=1$ case of the dilaton relation by explicit evaluation
    \be
        \<z\vp_I, \vp\>_W^1 = \<\vp\>_W.
    \ee
    In case $\vp= \vp_I$, both sides vanish by the degree selection. In case $\vp = \vp_P$ we use the LGS recursion 
    \be
    \begin{split}
        \<z\vp_I, \vp_P\>_W^1 &= \frac{d}{d\ve}\Big|_{\ve} \<z\vp_I\>^1_{W+\ve \vp_P}+\<C_W(z\vp_I, \vp_P)\>^1_W=\frac{d}{d\ve}\Big|_{\ve} \frac{1}{24}\sum_{p\in Crit(W)} 1 +\<\vp_P\>_W^1\\
        &=\frac{d}{d\ve}\Big|_{\ve} \frac{1}{12} +  \<\vp_P\>_W^1 =  \<\vp_P\>_W^1.
    \end{split}    
    \ee
    For $n>1$, we have two cases: all descendants have the same level, or descendants have different levels. If all descendants have level one, then both sides in (\ref{eq_LGS_dilaton}) are extremal correlators. We use the extremal correlation formula (\ref{def_LGS_extremal}) to evaluate both sides. Note that the residues are identical on both sides, while the dilaton relation follows from 
    \be
        \<\tau_1^{n+1}\>_1 = n \<\tau_1^n\>_1.
    \ee
    If the descendant levels are unequal, the correlation functions on both sides are non-vanishing if at least one level-zero descendant exists. We rearrange the observables so that $m_n=0$ and use the LGS recursion for $\vp_n$. We apply the LGS recursion to the following expression
    \be\nn
    \begin{split}
        \<&z^{m_1}\vp_1,\ldots, z^{m_{n-1}}\vp_{n-1},z\vp_I,\vp_{n}\>^1_{W}-(n-1)\<z^{m_1}\vp_1,\ldots, z^{m_{n-1}}\vp_{n-1},\vp_n\>^1_{W}\\
        &=\frac{d}{d\ve}\Big|_{\ve=0}\<z^{m_1}\vp_1,\ldots,z^{m_{n-1}}\vp_{n-1},z\vp_I\>^1_{W+\ve\vp_{n}}-(n-1)\frac{d}{d\ve}\Big|_{\e=0}\< z^{m_1}\vp_1,\ldots,z^{m_{n-1}}\vp_{n-1}\>^1_{W+\ve \vp_{n}}\\  
        &+\<z^{m_1}\vp_1,\ldots,z^{m_{n-1}}\vp_{n-1},C_W(z\vp_I,\vp_{n})\>^1_{W}+\sum_{k=1}^{n-1}\<z^{m_1}\vp_1,\ldots,C_{W}(z^{m_k}\vp_k,\vp_{n}),\ldots,z^{m_{n-1}}\vp_{n-1},z\vp_I\>_{W}\\
        &\qquad\qquad-(n-1)\sum_{k=1}^{n-1}\< z^{m_1}\vp_1,\ldots,C_{W}(z^{m_k}\vp_k,\vp_{n}),\ldots,z^{m_{n-1}}\vp_{n-1}\>^1_{W}\\
        &=\<z^{m_1}\vp_1,z^{m_2}\vp_2,\ldots, z^{m_{n-1}}\vp_{n-1},\vp_{n}\>^1_{W}.
    \end{split}
    \ee
    The terms in the second line and the sums cancel by the induction assumption. Note that the equality 
    \be\nn
        \< z^{m_1}\vp_1,\ldots,z^{m_{n-1}}\vp_{n-1}, z\vp_I\>^1_{W+\ve \vp_{n}}=(n-1)\< z^{m_1}\vp_1,\ldots,z^{m_{n-1}}\vp_{n-1}\>_{W+\ve\vp_{n}}^1
    \ee
    requires $n$-point equality for the superpotential deformed by the good section. For the mirror superpotential, the space of deformations is 2-dimensional. Each deformation is either along the trivial class $\vp_{I}$ or the point class $\vp_{P}$. The deformation along the trivial class $\vp_{n} = \vp_{I}$ preserves the LGS correlation functions. The deformation along the point class $\vp_n = \vp_P$ is equivalent to shifting $q$ in the mirror superpotential under the induction assumption. Hence, we arrive at equality 
    \be\label{eq_simpl_lgs_rec_dilaton_rel}
    \begin{split}
        \<&z^{m_1}\vp_1,\ldots, z^{m_{n-1}}\vp_{n-1},z\vp_I,\vp_{n}\>^1_{W}-(n-1) \<z^{m_1}\vp_1,\ldots, z^{m_{n-1}}\vp_{n-1}, \vp_n\>^1_{W}\\
        &=\<z^{m_1}\vp_1,z^{m_2}\vp_2,\ldots, z^{m_{n-1}}\vp_{n-1},\vp_{n}\>^1_{W}.
    \end{split}
    \ee
    The equality (\ref{eq_simpl_lgs_rec_dilaton_rel}) implies the dilaton relation (\ref{eq_LGS_dilaton}) for the $(n+1)$-point correlation function.
\end{proof}

\subsection{LGS TRR at genus one}
The key result to prove mirror symmetry is the LGS version of the topological recursion relation (LGS TRR).
\begin{Theorem} {\bf (LGS TRR)}
    For $n>0$ and integers $m_1\geq 1$ and $m_k>0$ in the LGS theory with superpotential $W$ and good section $\vp_k$
    \be\label{eq_genus_one_LGS_TRR}
    \begin{split}
        \<z^{m_1}\vp_1&,\ldots,z^{m_n}\vp_n\>^1_W = \frac{1}{24}\<z^{m_1-1}\vp_1,z^{m_2}\vp_2,\ldots, z^{m_n}\vp_n,\vp_a,\vp_b \>^0_W \eta^{ab} \\
         &+\sum_{S_1\neq \emptyset, S_1\sqcup S_2 = \{2,\ldots,n\}}\eta^{ab} \<z^{m_1-1}\vp_1,\{z^{m_\a}\vp_{\a}\}_{\a \in S_1},\vp_a\>_W^0
         \<\vp_b,\{z^{m_\b}\vp_\b\}_{\b\in S_2}\>^1_W.
    \end{split}     
    \ee
\end{Theorem}
To prove the theorem, we introduce several lemmas
\begin{Lemma}\label{lemma_W_double_prime} 
    For holomorphic functions $f_1, f_2$, good section basis $\vp_a$ and metric (\ref{eq_intersection_matrix})
    \be
    \begin{split}
        & \oint \frac{\vp_a\vp_b}{W'} = \eta_{ab},\\
        &\eta^{ab}\oint W''\frac{f_1 \vp_a}{W'}\oint \frac{f_2 \vp_b}{W'} = \oint W''\frac{f_1 f_2}{W'},  \\
        &\eta^{ab} \oint \frac{f_1\vp_a\vp_b}{W'}=\eta^{ab} \sum_{p\in Crit(W)} \frac{f_1\vp_a\vp_b}{W''}(p)=\oint \frac{f_1 W''}{W'} =\sum_{p\in Crit(W)}f_1(p).
    \end{split}
    \ee
\end{Lemma}
\begin{proof}
    The first equality is a straightforward evaluation of the residue integral for the basis $\vp_I, \vp_P$. For the second equality, we expand $f_2$ into a linear combination of a good section and an element from the Jacobi ideal 
    \be
        f_2 = f_2^\beta \vp_\beta + v_2W'. 
    \ee
    The $W'$-terms do not change the residue integral, so we can add and remove them. Hence, 
    \be
    \begin{split}
        &\eta^{ab}\oint W''\frac{f_1 \vp_a}{W'}\oint \frac{f_2 \vp_b}{W'}  =\eta^{ab}\oint W''\frac{f_1 \vp_a}{W'}\oint \frac{ (f_2^\beta \vp_\b+v_2W')\vp_b}{W'} 
        =\eta^{ab}\oint W''\frac{f_1 \vp_a}{W'}f_2^\beta\oint \frac{\vp_\b\vp_b}{W'} \\
        &=\eta^{ab}\oint W''\frac{f_1 \vp_a}{W'}f_2^\beta \eta_{b\beta} = f_2^a\oint W''\frac{f_1 \vp_a}{W'} =\oint W''\frac{f_1 (f_2^a\vp_a+v_2W')}{W'} 
        =\oint W''\frac{f_1f_2}{W'}. 
    \end{split}    
    \ee
    We prove the third equality by evaluating both sides for the mirror superpotential for $\mathbb{P}^1$.
    \be
    \begin{split}
        &\eta^{ab} \sum_{Crit(W)} \frac{\vp_I\vp_a\vp_p}{W''}(p) = 2 \sum_{Crit(W)} \frac{\vp_I\vp_I\vp_P}{W''}(p)=2=\sum_{crit(W)}\vp_I(p),\\
        &\eta^{ab} \sum_{Crit(W)} \frac{\vp_P\vp_a\vp_p}{W''}(p) = 2 \sum_{Crit(W)} \frac{\vp_I\vp_P\vp_P}{W''}(p)=0=\sum_{crit(W)}\vp_P(p).
    \end{split}    
    \ee  
\end{proof}
\begin{Lemma}\label{lemma_genus_one_TRR_extremal}
    LGS TRR (\ref{eq_genus_one_LGS_TRR}) holds for the over-extremal and extremal correlators.
\end{Lemma}
\begin{proof}
    The lhs is (over)extremal if the total descendant level is (equal to or) greater than the number of observables. Namely, we introduce the excess $M-n+3$ of the descendant level 
    for genus zero and $M-n$ for genus one, where we used the total descendant level
    \be
        M= \sum_{i=1}^n m_i.
    \ee
    The excess in the first term in the rhs of (\ref{eq_genus_one_LGS_TRR})
    \be
        0+0+m_1-1+ \sum_{i=2}^n m_i - (n+2)+3 = M-n.
    \ee
    The total excess of the descendant levels in the second line terms in the rhs of (\ref{eq_genus_one_LGS_TRR})
    \be
    \begin{split}
        &\left(m_1-1 + \sum_{\a \in S_1} m_\a- (|S_1|+2)+3\right) + \left(\sum_{\b\in S_2} m_\b -(|S_2|+1) \right) \\
        &= m_1+\sum_{\a\in S} m_\a -|S|-1 = M-n.
    \end{split}
    \ee
    Hence, if the lhs of (\ref{eq_genus_one_LGS_TRR}) is over-extremal, so are all terms in the rhs of (\ref{eq_genus_one_LGS_TRR}). Then both sides vanish, and the LGS TRR holds for a trivial reason. Similarly, if the lhs of (\ref{eq_genus_one_LGS_TRR}) is extremal, then the first term in the rhs is also (genus-zero) extremal. Hence, we can use an explicit formula for the extremal correlators to complete the proof. The lhs evaluates to
    \be\nn
        \<z^{m_1} \vp, z^{m_2}\vp_2, \ldots, z^{m_n}\vp_n\>^1_W = \<\tau_{m_1}\cdots \tau_{m_n}\>_1 \oint \frac{W''}{W'} \vp_1\vp_2\cdots \vp_n(p).
    \ee  
    The rhs evaluates to 
    \be\nn
    \begin{split}
         &\<z^{m_1-1} \vp, z^{m_2}\vp_2, \ldots, z^{m_n}\vp_n, \vp_a, \vp_b \>^0_W \eta^{ab} =\<\tau_{m_1-1}\tau_{m_2}\cdots \tau_{m_n}\tau_0\tau_0\>_0
         \eta^{ab}\oint \frac{\vp_a\vp_b\vp_1\cdots \vp_n}{W'}.\\
         & \sum_{S_1\neq \emptyset, S_1\sqcup S_2 = \{2,\ldots,n\}}\eta^{ab} \<z^{m_1-1}\vp_1,\{z^{m_\a}\vp_{\a}\}_{\a \in S_1},\vp_a\>_W^0
         \<\{z^{m_\b}\vp_\b\}_{\b\in S_2},\vp_b\>^1_W\\
         &=\sum_{S_1\neq \emptyset,S_1\sqcup S_2 = \{2,\ldots,n\}}\<\tau_0\tau_{m_1-1}\prod_{\a\in S_1}\tau_{m_\a}\>_0\<\tau_0\prod_{\b\in S_2}\tau_{m_\b}\>_1 \eta^{ab}
         \oint \frac{\vp_1 \vp_a}{W'} \prod_{\a\in S_1}\vp_{\a} \oint \frac{\vp_b}{W'}W''\prod_{\b\in S_2}\vp_{\b}.
    \end{split}     
    \ee
    By Lemma \ref{lemma_W_double_prime}, the contributions from the good sections are identical, so we just need to prove the equality of coefficients
    \be\label{eq_trr_genus_one_point}
    \begin{split}
        \<\tau_{m_1}\tau_{m_2}\cdots \tau_{m_n}\>_1& =\frac{1}{24}\<\tau_{m_1-1}\tau_0\tau_0\tau_{m_2}\cdots \tau_{m_n}\>_0\\
        &\qquad+\sum_{S_1\neq \emptyset, S_1\sqcup S_2 = \{2,\ldots,n\}}\<\tau_0\tau_{m_1-1}\prod_{\a\in S_1}\tau_{m_\a}\>_0\<\tau_0\prod_{\b\in S_2}\tau_{m_\b}\>_1.
    \end{split}    
    \ee
    The relation (\ref{eq_trr_genus_one_point}) is a topological recursion relation for the integrals of psi-classes over the moduli space of complex structures, as in Witten \cite{witten1990two}. 
\end{proof}

\begin{Theorem} {\bf (LGS TRR)} 
    LGS TRR (\ref{eq_genus_one_LGS_TRR}) holds for $n\geq 1, m_1>0$ and $m_k\geq 0$ and a collection of good sections $\vp_1,\ldots, \vp_n$.
\end{Theorem}
\begin{proof}     
    We have already proven the extreme and over-extreme cases in Lemma \ref{lemma_genus_one_TRR_extremal}. For the under-extremal case, we prove the LGS TRR by induction on $n$. The under-extremality implies that there is at least one observable with $m=0$. Assume $m_n=0$, use the LGS recursion for the observable $\vp_n$, and then use the LGS TRR for the $(n-1)$ points. The LGS recursion can produce $z^{m_1}\vp_1^\ve$ with lower powers of $z$, while the LGS TRR is only valid for non-zero $z$-powers. For the mirror superpotential for $\mathbb{P}^1$, the deformation contains only the $z^{m_1-1}$ term. Namely, 
    \be\label{eq_deform_lgs_descend}
        z^{m_1}\vp_1^\ve=z^{m_1}\vp_1+\ve z^{m_1}C_W(\vp_1,\vp_n)+\ve z^{m_1-1} S_W\pi_W(\vp_1\vp_n).
    \ee
    Hence, we need a separate treatment for the $S_W\pi_W(\vp_1\vp_n)$-term in the $m_1=1$ special case. For $m_1>1$, we have an equality
    \be\label{eq_lgs_recurstion_trr}
    \begin{split}
        &\<z^{m_1}\vp_1, z^{m_2}\vp_2, z^{m_3}\vp_3, \ldots, z^{m_{n-1}}\vp_{n-1},\vp_n\>^1_W  = \frac{d}{d\ve}\Big|_{\ve=0}\<z^{m_1}\vp^\ve_1, \ldots, z^{m_{n-1}}\vp^\ve_{n-1}\>^1_{W^\ve}\\
        & = \frac{1}{24}\frac{d}{d\ve}\Big|_{\ve=0} \<z^{m_1-1}\vp^\ve_1,z^{m_2}\vp^\ve_2,\ldots, z^{m_n}\vp^\ve_n,\vp^\ve_a,\vp^\ve_b \>^0_{W^\ve} \eta^{ab}\\
        & +\frac{d}{d\ve}\Big|_{\ve=0} \sum_{S_1\sqcup S_2 =  \{2,\ldots,n-1\}} \eta^{ab}\< \vp^\ve_a, z^{m_1-1}\vp^\ve_1, \{z^{m_\a}\vp^\ve_\a\}_{\a\in S_1}\>^0_{W^\ve} \< \vp^\ve_b,\{z^{m_\b}\vp^\ve_\b\}_{\b\in S_2}\>^1_{W^\ve}\\ 
        & = \frac{1}{24}\frac{d}{d\ve}\Big|_{\ve=0} \<z^{m_1-1}\vp^\ve_1,z^{m_2}\vp^\ve_2,\ldots, z^{m_n}\vp^\ve_n,\vp^\ve_a,\vp^\ve_b \>^0_{W^\ve} \eta^{ab}\\
        & +\sum \eta^{ab} \frac{d}{d\ve}\Big|_{\ve=0}\< \vp^\ve_a, z^{m_1-1}\vp^\ve_1, \{z^{m_\a}\vp^\ve_\a\}_{\a\in S_1}\>^0_{W^\ve} \< \vp_b,\{z^{m_\b}\vp_\b\}_{\b\in S_2}\>^1_W\\
        &\qquad+\sum \eta^{ab} \< \vp_a, z^{m_1-1}\vp_1, \{z^{m_\a}\vp_\a\}_{\a\in S_1}\>^0_W \frac{d}{d\ve}\Big|_{\ve=0}\< \vp_b,\{z^{m_\b}\vp^\ve_\b\}_{\b\in S_2}\>^1_{W^\ve}.
    \end{split}  
    \ee
    We dropped the arguments in the later summations since they are identical to the first sum. We need to compare our expression with the LGS recursion over $\vp_n$ of the rhs of the LGS TRR relation. The $\frac{1}{24}$-term 
    \be\label{eq_lgs_trr_recursion_24}
    \begin{split}
        \frac{1}{24}&\<z^{m_1-1}\vp_1,z^{m_2}\vp_2,\ldots, z^{m_n}\vp_n,\vp_a,\vp_b \>^0_W \eta^{ab}\\  
        &\qquad\qquad=\frac{1}{24}\frac{d}{d\ve}\Big|_{\ve=0} \<(z^{m_1-1}\vp_1)^\ve,z^{m_2}\vp^\ve_2,\ldots, z^{m_n}\vp^\ve_n,\vp^\ve_a,\vp^\ve_b \>^0_{W^\ve} \eta^{ab}.
    \end{split}    
    \ee
    Here, for $m_1>1$, we introduced
    \be
        (z^{m_1-1}\vp_1)^\ve = z^{m_1-1}\vp_1 +\ve z^{m_1-1}C_W(\vp_1,\vp_n)+\ve z^{m_1-2} S_W\pi_W(\vp_1\vp_n) = z^{m_1-1}\vp_1^\ve.  
    \ee
    The last equality implies that the two $\frac{1}{24}$ in (\ref{eq_lgs_recurstion_trr}) and (\ref{eq_lgs_trr_recursion_24}) are identical for $m_1>1$ as we anticipated.
    
    The LGS recursion for the term with the sum. Note that the $\vp_n$ may enter the sum in three different ways: it can be part of $S_1$, part of $S_2$, or a single observable in the first factor. Namely,
    \be\label{eq_lgs_trr_recursion_quadratic}
    \begin{split}
        &\sum_{S_1\sqcup S_2 = \{2,\ldots,n\}}\eta^{ab} \<z^{m_1-1}\vp_1,\{z^{m_\a}\vp_{\a}\}_{\a \in S_1},\vp_a\>_W^0\<\vp_b,\{z^{m_\b}\vp_\b\}_{\b\in S_2}\>^1_W\\
        &=\sum_{S_1\sqcup S_2 = \{2,\ldots,n-1\}}\eta^{ab} \<z^{m_1-1}\vp_1,\vp_n,\{z^{m_\a}\vp_{\a}\}_{\a \in S_1},\vp_a\>_W^0\<\vp_b,\{z^{m_\b}\vp_\b\}_{\b\in S_2}\>^1_W\\
        &+\sum_{S_1\sqcup S_2 = \{2,\ldots,n-1\}}\eta^{ab} \<z^{m_1-1}\vp_1,\{z^{m_\a}\vp_{\a}\}_{\a \in S_1},\vp_a\>_W^0\<\vp_b,\vp_n,\{z^{m_\b}\vp_\b\}_{\b\in S_2}\>^1_W\\
        &+\eta^{ab} \<z^{m_1-1}\vp_1,\vp_n,\vp_a\>_W^0\<\vp_b,z^{m_2}\vp_2,\ldots, z^{m_n}\vp_n\>^1_W.
    \end{split} 
    \ee
    The last term in (\ref{eq_lgs_trr_recursion_quadratic}) vanishes for $m_1>1$, while the LGS recursion for the other two terms matches the corresponding terms in (\ref{eq_lgs_recurstion_trr}). Hence, the LGS TRR for $(n-1)$ observables for $m_1>0$ implies the LGS TRR for $n$-observables and $m_1>1$. 
    
    For $m_1=1$, the LGS recursion (\ref{eq_lgs_recurstion_trr}) contains a term with level-0 first observable, namely
    \be\nn
    \begin{split}
        &\frac{d}{d\ve}\Big|_{\ve=0}\<\ve S_W\pi_W(\vp_1\vp_n),z^{m_2}\vp_2, \ldots, z^{m_{n-1}}\vp^\ve_{n-1}\>^1_{W^\ve}=\< S_W\pi_W(\vp_1\vp_n), z^{m_2}\vp_2, \ldots, z^{m_{n-1}}\vp_{n-1}\>^1_{W}.
    \end{split}
    \ee
    This extra term cancels with the last term in (\ref{eq_lgs_trr_recursion_quadratic}) due to the relation 
    \be
        \eta^{ab} \vp_b \<\vp_1,\vp_n,\vp_a\>_W^0 = S_W\pi_W(\vp_1\vp_n).
    \ee
     Hence, the LGS TRR for $n-1$ points and the LGS recursion imply the LGS TRR for $n$ points for all values of $m_1$.
\end{proof}

\section{Mirror at genus one}
In our works \cite{Losev:2022tzr,losev2023tropical} on tropical mirror symmetry, we proved the mirror relation between the GW theory on toric space $X$ and the correlators in mirror LGS theory. We can summarize the relation in the form
\begin{Proposition} \label{prop_genus_1_mirror_no_desc}{\bf Genus-1 mirror without descendants} 
    For $g=0,1$ and $n+k\geq 3-2g$ 
    \be
        \<\underbrace{I\cdots I}_k\cdot \underbrace{P\cdots P}_n\>^{\mathbb{P}^1}_{g} = \< \underbrace{\vp_I,\ldots,\vp_I}_k,\underbrace{\vp_P,\ldots, \vp_P}_n\>^g_W.
    \ee
\end{Proposition}
\begin{proof} 
    We proved the $g=0$ case in an earlier paper \cite{Lysov:2025nmn}, so we will focus on the $g=1$ case. The proof uses the explicit evaluation of both sides. Lemma \ref{lemma_GW_genus_one_no_descend} implies that the invariant is nonvanishing only for $k=0$ and $n=1$. The LGS puncture relation (\ref{eq_LGS_puncture}) implies that the LGS correlator vanishes for $k>0$ and $n\geq 1$. We use the divisor relation to evaluate the GW invariant for $k=0$. For $n>1$ the LGS divisor relation (\ref{eq_LGS_divisor}) implies
    \be\label{eq_n_point_lgs_points}
        \<\underbrace{\vp_P,\ldots, \vp_P}_n\>^1_W = (q\p_q)^{n-1} \< \vp_P\>^1_W = (q\p_q)^{n-1} \left(-\frac{1}{24}\right) = 0.
    \ee
    The final check is the $k=0, n=1$ and $k=1, n=0$ correlation functions. The GW expressions match the LGS expressions via 
    \be\nn
        \<I\>^{\mathbb{P}^1}_1 =0 = \<\vp_I\>^1_W,\;\; \<P\>^{\mathbb{P}^1}_1 =-\frac{1}{24} = \<\vp_P\>^1_W.
    \ee
\end{proof}
The proof of the theorem by an explicit evaluation shows that the mirror symmetry for $\mathbb{P}^1$ is extremely simple. However, in the presence of gravitational descendants, the GW invariants are more complicated, and the proof of the mirror symmetry is far from trivial.

\subsection{Mirror map for descendants}
We introduce the following elements of the good section
\be\label{eq_KM_coeffcients}
\begin{split}
    C_{2k}(P) &= \a_{2k} q^k \vp_P  = \frac{q^k}{k!^2} \vp_P,\;\;\; C_{2k+1}(P) = \a_{2k+1} q^{k+1}\vp_I =\frac{q^{k+1}}{k! (k+1)!}\vp_I;\\
    C_{2k}(I) &= \b_{2k} q^k \vp_I = \frac{q^k}{k!^2}(1-2k H_k)\vp_I,\;\; C_{2k+1}(I) = \b_{2k+1} q^{k}\vp_P =-\frac{2q^k}{k!^2}H_k \vp_P .
\end{split}
\ee
\begin{Definition}{\bf (Kontsevich-Manin map)}
    For $m\geq 0$ and cycle $\g$ the {\it mirror descendant} is the LGS observable
    \be\label{eq_KM_map}
        \Phi_{m}(\g) = \sum_{k=0}^m z^{k}C_{m-k}(\g) = C_m(\g) +z \Phi_{m-1}(\g).
    \ee
\end{Definition}
Kontsevich and Manin in \cite{kontsevich1998relations} instead of (\ref{eq_KM_map}) used a different formula
\be\label{eq_KM_map_2pt_funct}
    \Phi_{m}(\g) =\< \tau_{m-1}(\g) \g_a\> \eta^{ab} \vp_b +z \Phi_{m-1}(\g),\;\; \Phi_0(\g) = \vp_\g.
\ee
However, the two expressions (\ref{eq_KM_map}) and (\ref{eq_KM_map_2pt_funct}) coincide when we use the genus-zero 2-point functions for the GW theory on $\mathbb{P}^1$.

The first several descendants of a point class are
\be\label{eq_low_point_desc_mirr_map}
\begin{split}
    \Phi_0(P) &= C_0(P) = \vp_P = qe^{-iY},\\
    \Phi_{1}(P) &= C_1(P) + z \Phi_{0}(P) =  q \vp_I+z\vp_P = q + z qe^{-iY},\\
    \Phi_{2}(P) &= C_2(P) + z \Phi_{1}(P) =  q\vp_P+zq\vp_I+z^2 \vp_P,\\
    \Phi_{3}(P) &= C_3(P) + z \Phi_{2}(P) = \frac12 q^2\vp_I +zq\vp_P+z^2q+z^3 \vp_P.
\end{split}
\ee
The first several descendants of the identity class are
\be\label{eq_low_identity_desc_mirr_map}
\begin{split}
    \Phi_0(I) &=C_0(I) =  \vp_I =1,\\
    \Phi_1(I) &=C_1(I) + z \Phi_{0}(I)  =  z\vp_I =z,\\
    \Phi_2(I) &= C_2(I) + z \Phi_{1}(I)  = -q\vp_I +z^2 \vp_I = -q+z^2,\\
    \Phi_3(I) &= C_3(I) + z \Phi_{2}(I) = -2q\vp_P -qz\vp_I +z^3\vp_I = -2q^2 e^{-iY} -qz+z^3.
\end{split}
\ee

\subsection{Mirror for correlation functions}
Our main result is that the LGS correlation functions equal the GW invariants.
\begin{Theorem} {\bf (Mirror relation for genus-one)} \label{thm_main}
    The correlation functions in the genus $g=0,1$ descendant GW theory on $\mathbb{P}^1$ match the LGS correlation functions in the mirror LGS theory of the corresponding mirror LGS descendant invariants. Namely, for $g=0,1$, $n\geq 3-2g$, $m_n\geq 0$ and $\g_n \in H^\ast (\mathbb{P}^1)$
    \be
        \<\tau_{m_1}(\g_1)\cdots \tau_{m_n}(\g_n)\>_g = \< \Phi_{m_1}(\g_1),\ldots, \Phi_{m_n}(\g_n)\>^g_W.
    \ee
\end{Theorem}
\begin{proof} 
    The Dubrovin reconstruction Theorem \ref{thm_reconstruction} implies that the correlation functions of descendants in genus-one GW theory are uniquely determined by the TRR. We must also include the genus-zero GW invariants, genus-one GW invariants without descendants, the puncture relation, and the divisor relations. In our earlier paper \cite{Lysov:2025nmn} we showed that the genus-zero GW invariants on $\mathbb{P}^1$ match the genus-zero LGS correlation functions of the corresponding good sections. Proposition \ref{prop_genus_1_mirror_no_desc} establishes equality for no-descendant GW invariants to the corresponding genus-one LGS correlation functions. Hence, we only need to show that the LGS correlation functions of the mirror observables satisfy the GW puncture, divisor, and TRR relations. 
\end{proof}

\subsection{Puncture and divisor relations}
\begin{Proposition}
    The mirror map and the LGS puncture relation imply the GW puncture relation. Namely, for $g=0,1$ and $n\geq 3-2g$ and $m_k\geq 0$
    \be
        \< \Phi_{m_1}(\g_1),\ldots,\Phi_{m_n}(\g_n),\vp_I\>^g_W = \sum_{k=1}^n \< \Phi_{m_1}(\g_1),\ldots,\Phi_{m_k-1}(\gamma_k), \ldots,\Phi_{m_n}(\g_n)\>^g_W.
    \ee
\end{Proposition}
\begin{proof}
    We have already proven the genus-zero case in our earlier work \cite{Lysov:2025nmn}, so we will focus on the genus-one case. For $n\geq 1$, we use the LGS recursion with respect to the deformation by an identity observable 
    \be
        \< \Phi_{m_1}(\g_1),\ldots,\Phi_{m_n}(\g_n),\vp_I\>^1_W=\sum_{k=1}^n \<\Phi_{m_1}(\g_1),\ldots,C_W(\Phi_{m_k}(\gamma_k),\vp_I),\ldots,\Phi_{m_n}(\g_n)\>^1_W.
    \ee
    The contact terms evaluate to
    \be
        C_W(\Phi_{m}(\gamma),\vp_I) = C_W(C_m(\gamma) +z\Phi_{m-1}(\gamma),\vp_I) = \Phi_{m-1}(\gamma).
    \ee
\end{proof}
\begin{Proposition}
    The mirror map and the LGS divisor relation imply the GW divisor relation. Namely, for $g=0,1$, $n\geq 3-2g$ and $m_k\geq 0$
    \be
    \begin{split}
        \< &\Phi_{m_1}(\g_1),\ldots,\Phi_{m_n}(\g_n), \varphi_P\>^g_W=q\frac{d}{dq}\<\Phi_{m_1}(\g_1),\ldots,\Phi_{m_n}(\g_n)\>^g_W \\
        &+ \sum_{k=1}^n\<\Phi_{m_1}(\g_1),\ldots,\Phi_{m_k-1}(\g_k\wedge\g_P)\ldots,\Phi_{m_n}(\g_n)\>^g_W.
    \end{split}
    \ee
\end{Proposition}
\begin{proof}
    We have already proven the genus-zero case in our earlier work \cite{Lysov:2025nmn}. Note that the Kontsevich-Manin map (\ref{eq_KM_map}) and the LGS divisor relation are the same in genus 1 and genus 0. Hence, we can use the same proof.
\end{proof}

\subsection{Topological recursion relation at genus one}
Recall that the GW TRR \ref{eq_GW_trr_genus_one} allows $S_1=\emptyset$, while the LGS TRR (\ref{eq_genus_one_LGS_TRR}) does not, so we need to include the $S_1=\emptyset$ term carefully.  
\begin{Proposition} 
    For $n\geq 1$, $m_1>0$ and $m_k\geq 0$
    \be\label{eq_gw_trr_via_lgs_trr}
    \begin{split}
        \<&\Phi_{m_1}(\gamma_1),\ldots,\Phi_{m_n}(\gamma_n)\>^1_W=\<\tau_{m_1-1}(\g_1),\g_a\>_0 \eta^{ab}\<\vp_b,\Phi_{m_2}(\gamma_2),\ldots,\Phi_{m_n}(\gamma_n)\>^1_W\\
        &\qquad\qquad+\sum \<\vp_a,\Phi_{m_1-1}(\gamma_1),\{\Phi_{m_\a}(\gamma_\a)\}_{\a\in S_1}\>^0_W \eta^{ab} \<\vp_b,\{\Phi_{m_\b}(\gamma_\b)\}_{\b\in S_2}\>^1_W\\
        &\qquad\qquad+\frac{1}{24} \eta^{ab} \<\Phi_{m_1-1}(\gamma_1),\ldots,\Phi_{m_n}(\gamma_n),\vp_a,\vp_b\>^0_W.
    \end{split}
    \ee
    The sum is taken over possible subsets $S_1\neq \emptyset$ and $S_2$ such that $S_1\sqcup S_2 = \{2,\ldots,n\}$.
\end{Proposition}
\begin{proof}
    LGS correlation function is linear in all arguments. However, the key difference between the LGS TRR and GW TRR is the presence of the 2-point function in the GW TRR (\ref{eq_GW_trr_genus_one}). Let us use the KM map to rewrite the lhs of (\ref{eq_gw_trr_via_lgs_trr}) 
    \be\label{eq_proof_GW_trr_from_LGS}
    \begin{split}
        \<\Phi_{m_1}(\gamma_1),\Phi_{m_2}(\gamma_2),\ldots, &\Phi_{m_n}(\gamma_n)\>^1_W =\<z\Phi_{m_1-1}(\gamma_1), \Phi_{m_2}(\gamma_2),\ldots,\Phi_{m_n}(\gamma_n)\>^1_W \\
        &+\<C_{m_1}(\gamma_1),\Phi_{m_2}(\gamma_2),\ldots,\Phi_{m_n}(\gamma_n)\>^1_W .
    \end{split}
    \ee
    We use the LGS TRR to express the first term in (\ref{eq_proof_GW_trr_from_LGS})
    \be
    \begin{split}
        \<&z\Phi_{m_1-1}(\gamma_1),\Phi_{m_2}(\gamma_2),\ldots,\Phi_{m_n}(\gamma_n)\>^1_W\\
        &=\sum \<\vp_a,\Phi_{m_1-1}(\gamma_1),\{\Phi_{m_\a}(\gamma_\a)\}_{\a\in S_1}\>^0_W \eta^{ab} \<\vp_b,\{\Phi_{m_\b}(\gamma_\b)\}_{\b\in S_2}\>^1_W\\
        &\qquad\qquad+\frac{1}{24} \eta^{ab} \<\Phi_{m_1-1}(\gamma_1),\Phi_{m_2}(\gamma_2),\ldots,\Phi_{m_n}(\gamma_n),\vp_a,\vp_b\>^0_W.
    \end{split}
    \ee
    The sum is taken over possible subsets $S_1\neq \emptyset$ and $S_2$ such that $S_1\sqcup S_2 = \{2,\ldots,n\}$. 

    We use the KM map (\ref{eq_KM_map_2pt_funct}) to rewrite the second term in (\ref{eq_proof_GW_trr_from_LGS}) in a form that matches the first term in (\ref{eq_gw_trr_via_lgs_trr})
    \be\nn
        \<C_{m_1} (\gamma_1), \Phi_{m_2}(\gamma_2),\ldots,\Phi_{m_n}(\gamma_n)\>^1_W 
        =\<\tau_{m_1-1}(\g_1), \g_a\>_0 \eta^{ab}\< \vp_b, \Phi_{m_2}(\gamma_2),\ldots,\Phi_{m_n}(\gamma_n) \>^1_W.
    \ee
\end{proof}

\section{Examples and applications}
This section describes several examples and general applications of the mirror LGS description of GW invariants, such as the polynomiality property.
\subsection{1-point functions at genus one}
There is a special 1-point function in GW theory in genus one on $X$
\be
    \<\tau_1(I)\>_1 = \frac{1}{24} \chi(X).
\ee
Moreover, the toric space $X$ has only even cohomology, so 
\be
    \chi(X) = \dim H^{\ast} (X).
\ee
The KM map for $\tau_1(1)$ is universal for all $X$, namely $\Phi_1(I) = z \vp_I = z \cdot 1$. Hence, the corresponding LGS correlator is a purely extremal one
\be
    \<\Phi_1(I)\>_W^1 = \<z\vp_I\>_W^1 = \<\tau_1\>_1 \sum_{p \in Crit(W)} \vp_I(p) = \frac{1}{24} \dim J_W = \frac{1}{24}\dim H^\ast(X).
\ee
\begin{Proposition} The LGS mirror holds for genus-one 1-point functions. Namely,
\be
    \<\tau_{m}(\g)\>_1 = \<\Phi_{m}(\g)\>_W^1,\;\; \g \in H^\ast (\mathbb{P}^1).  
\ee
\end{Proposition}
\begin{proof} Let us start with writing relations in components
\be
\begin{split}
    \<\tau_{2n}(P)\>_1& =\frac{2n-1}{24 n!^2}q^n =  \<\Phi_{2n}(P)\>_W^1, \\
    \<\tau_{2n+1}(P)\>_1 &=0 =  \<\Phi_{2n+1}(P)\>_W^1,\\
    \<\tau_{2n}(I)\>_1 &=0 =  \<\Phi_{2n}(I)\>_W^1,\\
    \<\tau_{2n+1}(I)\>_1& =\frac{1}{12 n!^2}q^n (1+(1-2n)H_n) =  \<\Phi_{2n+1}(I)\>_W^1.
\end{split}
\ee
We use the leading-order expansion for the mirror map of descendants 
\be
\begin{split}
    \Phi_{2n}(P) &= \frac{q^n}{n!^2}\vp_P + z \frac{q^n}{n! (n-1)!}\vp_I + \cO(z^2),\\
    \Phi_{2n+1}(P) &= \frac{q^{n+1}}{n!(n+1)!}\vp_I + z \frac{q^n}{n!^2}\vp_P + \cO(z^2),\\
    \Phi_{2n}(I) &= \frac{q^n}{n!^2}(-2n H_n +1)\vp_I + z \frac{q^{n-1}}{(n-1)!^2}(-2H_{n-1})\vp_P + \cO(z^2),\\
    \Phi_{2n+1}(I)& = \frac{q^n}{n!^2}(-2H_n)\vp_P + z\frac{q^n}{n!^2}(-2n H_n +1)\vp_I+\cO(z^2).
\end{split}
\ee
The corresponding correlation functions are 
\be
\begin{split}
    \<\Phi_{2n}(P)\>_W^1 &=\frac{q^n}{n!^2} \left(\<\vp_P\>_W^1+n\<z\vp_I\>_W^1 \right)=\frac{q^n}{n!^2} \left(-\frac{1}{24}+n\frac{1}{12}\right)= \frac{2n-1}{24 n!^2}q^n\\
    \<\Phi_{2n+1}(P)\>_W^1 &=\frac{q^n}{n!(n+1)!} \left( \<\vp_I\>_W^1 + (n+1) \<z\vp_P\>_W^1 \right) = \frac{q^n}{n!(n+1)!} (0+0) = 0,\\
    \<\Phi_{2n}(I)\>_W^1 &=(...) \left( \<\vp_I\>_W^1 + (...) \<z\vp_P\>_W^1 \right) = (...) (0+0) = 0,\\
    \<\Phi_{2n+1}(I)\>_W^1 & = \frac{q^n}{n!^2} (-2H_n \<\vp_P\>_W^1+(-2n H_n +1)\<z\vp_I\>_W^1)\\
&=\frac{q^n}{n!^2} \left( 2H_n\frac{1}{24} + (-2n H_n +1) \frac{1}{12} \right) =\frac{q^n}{12\cdot n!^2}(1+(1-2n)H_n).
\end{split}
\ee
\end{proof}

\subsection{Two-point functions of the point descendants at genus one}
The simplest 2-point function from (\ref{eq_GW_2pt_NS})
\be\nn
    \<\tau_1(P) \tau_1(P)\>_{1}  =\frac{1\cdot 1}{24\; 1!^2\; 1!^2} (2\cdot 1^2+2\cdot 1^2+2\cdot 1\cdot 1-3\cdot 1-3\cdot 1)=0.
\ee
The LGS computation 
\be\nn
\begin{split}
    \<\Phi_{1}(P),\Phi_{1}(P)\>^1_W &= q^2\<\vp_I,\vp_I\>^1_W  +2q\<\vp_I,z\vp_P\>^1_W +\<z\vp_P,z\vp_P\>^1_W \\
    & =q^2\cdot 0 +2q\<\vp_P\>^1_W +\<\tau_1^2\> \oint \frac{\vp_P\vp_P W''}{W'} =2q \cdot \left(-\frac{1}{24}\right) +\frac{1}{24}\cdot 2q=0.
\end{split}
\ee
The following proposition gives the general case.
\begin{Proposition} LGS mirror holds for genus-one 2-point functions. Namely,
\be\nn
\begin{split}
    \<\tau_{2n}(P) \tau_{2m}(P)\>_1 &=\frac{2n^2+2m^2+2mn-n-m}{24\; n!^2\; m!^2}q^{m+n} = \<\Phi_{2n}(P), \Phi_{2m}(P)\>^1_W,\\
    \<\tau_{2n-1}(P) \tau_{2m-1}(P)\>_1 &= \frac{mn}{24\; n!^2\; m!^2} (2n^2+2m^2+2mn-3m-3n)q^{m+n-1}= \<\Phi_{2n-1}(P), \Phi_{2m-1}(P)\>^1_W.
\end{split}
\ee
\end{Proposition}
\begin{proof} We use the leading order expansion for the mirror map of descendants 
\be
\begin{split}
    \Phi_{2n}(P) &= \frac{q^n}{n!^2}\vp_P + z \frac{q^n}{n! (n-1)!}\vp_I + z^2\frac{q^{n-1}}{(n-1)!^2}\vp_P +\cO(z^3)=\\
    &=\frac{q^n}{n!^2} (\vp_P + nz \vp_I + n^2q^{-1}z^2 \vp_P),\\
    \Phi_{2n+1}(P) &= \frac{q^{n+1}}{n!(n+1)!}\vp_I + z \frac{q^n}{n!^2}\vp_P + \frac{q^n}{n!(n-1)!}\vp_I+\cO(z^3).
\end{split}
\ee
The even descendants' case 
\be\nn
\begin{split}
    \left(\frac{q^{n+m}}{m!^2 n!^2}\right)^{-1} &\<\Phi_{2n}(P), \Phi_{2m}(P)\>^1_W = \< \vp_P + nz \vp_I + n^2q^{-1}z^2 \vp_P, \vp_P + mz \vp_I + m^2q^{-1}z^2 \vp_P\>^1_W \\
    &=\< \vp_P, \vp_P\>^1+(n+m) \<z \vp_I, \vp_P\>^1+mn\<z\vp_I, z\vp_I\>^1+(n^2 +m^2) q^{-1}\<z^2 \vp_P, \vp_P\>^1\\
    &= 0+(n+m) \< \vp_P\>^1+mn\<\tau_1^2\>_1 \oint \frac{W''}{W'}\vp_I^2 +(n^2 +m^2) q^{-1}\<\tau_2\tau_0\> \oint \frac{W''}{W'}\vp_P^2\\
    &=-\frac{n+m}{24} +\frac{2nm}{24} + \frac{2(n^2+m^2)}{24}.
\end{split}
\ee
We dropped all over the extremal correlators in the second line. We used the $\<\vp_P,\vp_P\>^1_W = 0$ dilaton relation and the extremal correlator formula. The proof for the case with both descendant levels being odd is very similar. 
\end{proof}

\subsection{Multipoint functions of level-1 descendants at genus one}

Dubrovin-Yang \cite{dubrovin2019gromov} provides explicit expressions for genus-one correlation functions of powers of $\tau_1(P)$.
\be
\begin{split}
    \<\tau_1(P)^2\>_1 =0,\;\;\; \<\tau_1(P)^4\>_1 =\frac12q^2,\;\;\; \<\tau_1(P)^6\>_1 =40 q^3.
\end{split}
\ee
The descendant map is 
\be
    \Phi_1(P) = q \vp_I + z\vp_P.
\ee
The LGS correlation function for the $n=4$ case is
\be\nn
\begin{split}
    &\<\Phi_{1}(P)^{\otimes 4}\>^1_W =\<q\vp_I+z\vp_P,q\vp_I +z\vp_P,q\vp_I+z\vp_P,q\vp_I+z\vp_P\>^1_W =q^4\<\vp_I^{\otimes 4}\>^1_W+4q^3\<\vp_I^{\otimes 3},z\vp_P\>^1_W\\
    &+6q^2\<\vp_I,\vp_I,z\vp_P,z\vp_P\>^1_W+4q\<\vp_I,z\vp_P,z\vp_P,z\vp_P\>_W^1+\<(z\vp_P)^{\otimes 4}\>_W^1 \\
    &=0+4q^3\<\vp_I,\vp_I,\vp_P\>^1_W +6\cdot 2 q^2\<\vp_P,\vp_P\>^1_W + 4\cdot 3q\<\vp_P,z\vp_P,z\vp_P\>_W^1+\<\tau_1^4\>_1\oint \frac{W''\vp_P^4}{W'}  \\
    &= 0 +0 + 4\cdot 3 q \<\vp_P,z\vp_P,z\vp_P\>+3!\cdot \frac{1}{24}\cdot 2q^2=0+\frac{1}{2}q^2=\frac{1}{2}q^2.
\end{split}
\ee
We used the extreme four descendant terms in the second line. We used the puncture relation on lines 3 and 4. We used the divisor relation to evaluate 
\be\nn
\begin{split}
    \<\vp_P,z\vp_P,z\vp_P\>^1_W& = q\p_q \<z\vp_P,z\vp_P\>^1_W +2 \<S\pi(\vp_P\vp_P),z\vp_P\>^1_W\\
    & =q\p_q \frac{q}{12}+2q\<1,z\vp_P\>^1_W = \frac{q}{12}+2q \<\vp_P\>^1_W =\frac{q}{12}+2q\left(-\frac{1}{24}\right) = 0.
\end{split}
\ee
The LGS correlation function for the $n=6$ case is
\be\nn
\begin{split}
    &\<\Phi_{1}(P)^{\otimes 6})\>^1_W =  \<(q\vp_I +z\vp_P)^{\otimes 6}\>^1_W \\
    &=q^6\<\vp_I^{\otimes 6}\>^1_W  +\binom{6}{1}q^5\<\vp_I^{\otimes 5},z\vp_P\>^1_W +\binom{6}{2}q^4\<\vp_I^{\otimes 4},z\vp_P, z\vp_P\>^1_W 
    +\binom{6}{3}q^3\<\vp_I^{\otimes 3},(z\vp_P)^{\otimes 3}\>^1_W \\
    &\qquad+ \binom{6}{4}q^2\<\vp_I,\vp_I,(z\vp_P)^{\otimes 4}\>^1_W +\binom{6}{5}q\<\vp_I,(z\vp_P)^{\otimes 5}\>^1_W +\<(z\vp_P)^{\otimes 6}\>^1_W \\
    &=3!\binom{6}{3}q^3\<\vp_P^{\otimes 3}\>^1_W + 2!\binom{6}{4}q^2\<\vp_P^{\otimes 2},z\vp_P,z\vp_P\>^1_W +5\binom{6}{5}q\<\vp_P,(z\vp_P)^{\otimes 5}\>^1_W+\<\tau_1^6\>_1 \oint \frac{W''\vp_P^6}{W'} \\
    &=5\binom{6}{5}q\cdot q^2+q^3 5!\frac{1}{24} \cdot 2 = 30 q^3 + 10 q^3 = 40 q^3.
\end{split}
\ee
We used the puncture relation and the extremal correlator in the first simplification. We used the no descendant mirror to evaluate $\<\vp^{\otimes 3}_P\>_W^1 = 0$. We simplify the remaining two terms using the divisor relation 
\be\nn
\begin{split}
    \<\vp_P,(z\vp_P)^{\otimes 4}\>_W &= q\p_q \<(z\vp_P)^{\otimes 4}\>^1_W + 4q\<\vp_I,(z\vp_P)^{\otimes 3}\>^1_W\\
    &= q\p_q (\<\tau_1^4\> 2q^2) + 4q \<\vp_P,z\vp_P, z\vp_P\>_W = 4q^2 \cdot 3! \cdot \frac{1}{24} = q^2. 
\end{split}
\ee
and
\be\nn
\begin{split}
    \<\vp_P^{\otimes 2},(z\vp_P)^{\otimes 2}\>_W &= q\p_q \<\vp_P,z\vp_P, z\vp_P\>_W + 2q \<\vp_P,1, z\vp_P\>_W  \\
    &= q\p_q (0) + 2q \<\vp_P,\vp_P\>_W = 0. 
\end{split}
\ee

\subsection{Polynomiality}
The GW invariants in our examples have the structure of a simple polynomial in the descendant levels divided by a product of factorials. Norbury and Scott in \cite{norbury2014gromov} formalized this observation and even provided a proof based on the topological recursion presentation for the GW invariants. In our proof, we use the mirror LGS theory. 
\begin{Theorem} {\bf (Norbury-Scott)} 
    For $g=0,1$ and even $k$ the GW invariants of $\mathbb{P}^1$ are of the form
    \be
        \left\< \prod_{i=1}^k \tau_{2m_i-1}(P) \prod_{i=k+1}^n \tau_{2m_k} (P) \right\>_g = \frac{m_{1}\cdots m_k}{m_1!^2 \cdots m_n!^2} p^g_{n,k} (m_1, \ldots, m_n)\; q^{1-k/2+\sum m_i}.
    \ee
    Here $p^g_{n,k}$ is a polynomial of degree $3g-3+n$ in $m_i$, symmetric in the first $k$ and last $n-k$ variables, with top coefficient $c_\beta$ of $m_1^{\beta_1}\cdots m_n^{\beta_n}$, for $|\beta| = 3g-3+n\geq 0$ given by 
    \be\label{eq_moduli_space_int_ns_polynom}
        c_\beta=2^g \int_{\overline{\mathcal{M}_{g,n}}} \psi_1^{\beta_1}\cdots \psi_n^{\beta_n}.
    \ee
\end{Theorem}
\begin{proof} 
    We proved the genus-zero version in \cite{Lysov:2025nmn}, so here we will focus on the genus-one case.
    
    We rewrite the mirror map (\ref{eq_KM_map}) for point descendants in the following form
    \be\label{eq_polyn_mirror_map}
    \begin{split}
        m!^2 \;\Phi_{2m}(P)&=\vp_P \sum_{k=0}^m q^{m-k} z^{2k}P_{2k}(m)  + \vp_I \sum_{k=0}^{m-1} q^{m-k} z^{2k+1}P_{2k+1}(m),\\
        m!(m-1)!\;\Phi_{2m-1}(P)&=\vp_P \sum_{k=1}^{m} q^{m-k} z^{2k-1}Q_{2k-1}(m)  + \vp_I \sum_{k=0}^{m-1} q^{m-k} z^{2k}Q_{2k}(m). 
    \end{split}
    \ee
    In (\ref{eq_polyn_mirror_map}) we introduced polynomials 
    \be\label{eq_polynomials_desc}
    \begin{split}
        P_{2k}(m) &= \prod_{j=0}^{k-1} (m-j)^2,\;\; P_{2k+1}(m)= m \prod_{j=0}^{k-1} (m-j)(m-1-j),\\
        Q_{2k}(m)&  = \prod_{j=0}^{k-1} (m-j)(m-j-1),\;\; Q_{2k+1}(m) = m\prod_{j=0}^{k-1} (m-j-1)^2.
    \end{split}
    \ee
    Moreover, in the expansion (\ref{eq_polyn_mirror_map}), the degree of the polynomials matches the power of $z$, the LGS descendant level. By definition of the LGS correlation functions, the genus-one correlation functions of $n$ observables vanish when the total descendant level is $n+1$ or more. The total descendant level is the total $z$-degree, and the corresponding LGS correlator is multiplied by the polynomial of $m_j$ of the same degree. Hence, the maximal degree of the polynomial is $n$, identical to the theorem's prediction at $g=1$. 
    
    The top-degree polynomial contributions are multiplied by the extreme correlators. Moreover, the top-degree term in polynomials (\ref{eq_polynomials_desc}) 
    \be
    \begin{split}
        P_{a}(m) &= m^{a} + \cO(m^{a-1}),\;\;Q_{a}(m) = m^{a} + \cO(m^{a-1}).
    \end{split}
    \ee
    Hence, the leading monomials with $|\beta| = \b_1+\ldots +\b_n  = n$ are of the form  
    \be\label{eq_leading_pol_term_point_desc}
        m_1^{\b_1}\cdots m_n^{\beta_{n}}\< z^{\b_1} \vp_{\a_1},\ldots, z^{\beta_n} \vp_{\a_n}\>^1_W = m_1^{\b_1}\cdots m_n^{\beta_{n}} \<\tau_{\beta_1}\cdots\tau_{\beta_n}\>_1\oint \frac{W''\vp_{\a_1}\cdots \vp_{\a_n}}{W'}.
    \ee
    The labels $\a_k \in \{I, P\}$ are such that the residue integral of good sections is $2=2^1 =2^g$ times the appropriate power of $q$.
    The multinomial factor in (\ref{eq_leading_pol_term_point_desc}) is identical to the moduli space integral (\ref{eq_moduli_space_int_ns_polynom}).
\end{proof}

\section*{Acknowledgments}

The author is grateful to Andrey Losev for discussing the topics presented in this paper and for his comments and feedback.

\bibliography{LGS_genus_one_ref}{}
\bibliographystyle{utphys}

\end{document}